\documentclass[a4paper]{article} 

\usepackage{booktabs}
\usepackage{authblk}
\usepackage{natbib}
\usepackage{graphicx}
\usepackage[table,xcdraw]{xcolor}
\usepackage{geometry}
\usepackage{float}

\title{\textbf{Forma mentis networks reconstruct how Italian high schoolers and international STEM experts perceive teachers, students, scientists, and school}}

\author[1]{Massimo Stella}

\affil[1]{Complex Science Consulting, Via Amilcare Foscarini 2, Lecce, Italy. Email: massimo.stella@inbox.com
}

\begin{document}

\maketitle

% Abstract (Do not insert blank lines, i.e.,~\\) 
\begin{abstract}
This study investigates how students and researchers shape their knowledge and perception of educational topics. The mindset or \textit{forma mentis}
of 159 Italian high school students and of 59 international researchers in science, technology, engineering and maths (STEM) are reconstructed through forma mentis networks, i.e.,~cognitive~networks of concepts connected by free associations and enriched with sentiment labels. The layout of conceptual associations between positively/negatively/neutrally perceived concepts is informative on how people build their own mental constructs or beliefs about specific topics. 
Researchers displayed mixed positive/neutral mental representations of ``teacher'', ``student'' and, ``scientist''. Students' conceptual associations of ``scientist'' were highly positive and largely non-stereotypical, although links about the ``mad scientist'' stereotype persisted. Students perceived ``teacher'' as a complex figure, associated with positive aspects like mentoring/knowledge transmission but also to negative sides revolving around testing and grading. ``School'' elicited stronger differences between the two groups. In the students' mindset, ``school'' was surrounded by a negative \textit{emotional aura} or set of associations, indicating an \textit{anxious} perception of the school setting, mixing scholastic concepts, anxiety-eliciting words, STEM disciplines like maths and physics, and exam-related notions. Researchers' positive stance of ``school'' included concepts of fun, friendship, and personal growth instead. Along the perspective of Education Research, the above results are discussed as quantitative evidence for test- and STEM anxiety {co-occurring} in the way Italian students perceive education places and their actors. Detecting these patterns in student populations through forma mentis networks offers new, simple to gather yet detailed knowledge for future data-informed intervention policies and action research.    
\end{abstract}

% Keywords

\section{Introduction}

Increasingly more studies in Education Science suggest that a large number of students experience anxiety while in educational settings~\cite{mallow2006science,zeidner2010test,nunez2013effects,lehtamo2018connection,siew2019using}. Stress can arise from multiple sources. Among~science, technology, engineering and mathematics (STEM) subjects,  disciplines like physics~\cite{mallow2006science}, maths~\cite{nunez2013effects}, and statistics~\cite{siew2019using} are increasingly reported to elicit fearful emotional states of higher alertness across different educational systems and levels, from~primary school to college students~\cite{nunez2013effects,siew2019using}. This phenomenon was first detected by Mallow, who also coined the term ``science anxiety''~\cite{mallow2006science}, and~then related it to poor performance in science courses~\cite{mallow2006science,bodin2012role,nunez2013effects,siew2019using}. The~heightened state of arousal induced by anxiety ends up inhibiting students' concentration on the task at end, importantly impacting problem solving~\cite{bodin2012role} and even knowledge retention~\cite{lehtamo2018connection}. Anxiety can originate from several sources in a given educational setting. Teachers' behavior can greatly affect classroom settings, ultimately provoking alertness in students either through active~\cite{mallow2006science} (e.g.,~using assessments in unbalanced ways) or passive behavior (e.g.,~neglecting peer-learning or cooperative learning ~\cite{oludipe2010effect,van2018educate}). Notice that personal assessment in Education is a rather delicate matter, with~imminent examinations frequently inducing the so-called ``test anxiety''~\cite{cassady2002cognitive,zeidner2010test}, a~feeling of distress, over-arousal, and tension that is mostly detected before taking an exam or a test. Cassady and Johnson showed that moderate levels of anxiety correlated positively with better academic achievements whereas stronger levels of anxiety impacted negatively test performance~\cite{cassady2002cognitive}. Importantly, science anxiety can arise also from a lack of role models and a generally dry perception of a subject~\cite{mallow2006science}. Recurrent stereotypes of scientists in terms of ``intelligent-yet-boring'' individuals~\cite{finson1995development} or ``evil geniuses''~\cite{haynes2016whatever} provide a distorted picture of the nature of science and knowledge creation and prevent students from getting inspirational role models, which can ultimately help dealing with stress and improve personal academic achievement~\cite{rahm2002scientist}. {In psychology, stereotypes summarize common expectations about the main features possessed by members of a given group}~\cite{beilock2007stereotype}. {Negative stereotypes can affect self-judgement, reduce working memory performance, elicit anxiety and drastically hinder academic performance}~\cite{beilock2007stereotype}, {a~phenomenon known as \textit{stereotype threat} and recently identified also in STEM education}~\cite{shapiro2012role}. It is important to underline that the development and exacerbation of anxious moods from all the above sources have detrimental effects for students' performance within school settings and beyond~\cite{bodin2012role,nunez2013effects,siew2019using}, decreasing academic achievement~\cite{mallow2006science,lehtamo2018connection} and preventing students from pursuing STEM careers~\cite{valenti2016adolescents}. Within~this complex picture, it becomes of utmost importance to develop general and simple techniques quantifying and detecting anxiety in the students' perception of their classroom settings, of~STEM subjects, of potential stereotypes and, more in general, of~the whole educational~system.

Forma mentis networks (FMN) were recently introduced by Stella and colleagues~\cite{stella2019forma} and shown to be capable of detecting anxiety patterns through easily obtainable cognitive data. Building on these results, this paper adopts the framework of forma mentis networks for better understanding the perception toward educational topics and STEM of high school students in their final year of studies, with~a focus on the Italian educational system. Being almost on the verge of graduation, these students  completed their career through primary education and most of secondary education and are therefore representative of the learning outcomes of such systems. Mapping the students' mindset or \textit{forma mentis} (in Latin) can shed light on particular anxiety-related, negative or positive stances that should be better investigated or acted upon by Education researchers, teaching professionals and policy makers for improving students' experience and their academic impact in~STEM.

FMNs combine network science and cognitive science methods and data~\cite{stella2019forma,stella2019formaindiv} with the aim of stance detection, i.e.,~identifying whether an individual or a group is in favor or against a given topic. Differently from machine learning approaches, often solving stance detection through black-box analyses of textual data~\cite{mohammad2016semeval}, forma mentis networks produce a transparent representation of how individuals or populations of individuals perceive and associate knowledge about a specific topic. Through free associations~\cite{de2013better} and sentiment patterns~\cite{smeets2006effect}, forma mentis networks can reconstruct how knowledge is represented, structured, and perceived by~people.

This mental representation of a stance is grounded in years of research at the fringe of computer science, psychology, and linguistics. Empirical and theoretical research in the cognitive sciences identified these mental representations of knowledge as components of a way more complex system called \textit{mental lexicon}, a~repository of knowledge apt at information acquisition, processing, and use~\cite{doczi2019overview}. The~recent adoption of network science tools has shown how the large-scale, associative structure of word knowledge in the mental lexicon is highly informative of a wide variety of cognitive processes such as lexical processing~\cite{de2013better,van2015examining,vitevitch2018spoken,neergaard2019phonological}, learning and cognitive development~\cite{stella2017multiplex,stella2018distance,stella2018multiplex}, text~structuring and writing styles~\cite{amancio2015complex,dearruda2019paragraph,machicao2018authorship,stella2018bots}, creativity~\cite{kenett2019semantic,stella2019viability}, and expertise levels in specific domains~\cite{tyumeneva2017distinctive,siew2019using}. Analogously, forma mentis networks act as approximated reconstructions on the mental constructs built by individuals in their associative mental lexicon, representing their perceptions of the outer world~\cite{stella2019forma,stella2019formaindiv}.

In comparison to concept maps, which have been successfully used in Education Research as tools for representing knowledge dependencies between parts of a syllabus or learning expertise and identifying key concepts for achieving learning outcomes, cfr.~\cite{koponen2006generative,nousiainen2010concept,tyumeneva2017distinctive,koponen2014systemic,mika2019maths,koponen2017complex,subramaniam2019using,kinchin2019uncovering}, FMNs represent a different cognitive dimension, depending on how individuals' memory associates knowledge rather than just representing pre-requisites or conceptual dependencies between concepts. FMNs can therefore represent a wider range of knowledge structures, free from any definition and mixing semantic, syntactic, visual or even phonological associations between concepts~\cite{de2013better}. Furthermore, FMNs include also a sentiment or affective component~\cite{smeets2006effect} that is absent in concept maps and which enables a closer understanding of emotional perception and anxiety-eliciting. Analogously to concept maps~\cite{koponen2017complex}, forma mentis networks can also~identify changes in the perception or mindset of students over time~\cite{stella2019formaindiv}.

With such a cognitive methodology, this paper assumes an educational scope by investigating stances toward STEM, Education, learning environments, and professionals as represented by the aggregated mindset of 159 Italian high schools students and of 59 international researchers (reconstructed through the data gathered in~\cite{stella2019forma}). Differently from past approaches, this analysis focuses not on STEM subjects but rather on educational concepts like ``teacher'', ``school'', ``scientist'', ``student'', or ``learning''. Most of the focus is devoted to comparisons between students' and researchers' perception and also to detecting how anxiety patterns potentially arise across all the considered educational stances. The~manuscript ends with a discussion of the above results in relation to the relevant~literature.

\section{Methods}

This section briefly reports about the data used for the current analysis. No new data were generated for this study. {Notice that this work is based on the data provided by Stella and colleagues}~\cite{stella2019forma}, {which included cues, responses, and valence labels (positive/negative/neutral). The~dataset was anonymous and it did not include individual demographics of students and researchers but only overall fractions of female/male participants and average age, which are reported in the following.}

\subsection{Forma Mentis~Networks}

This manuscript uses the forma mentis networks released by Stella and colleagues and representative of 159 Italian high school students and of 59 international young researchers on complex systems~\cite{stella2019forma}. The~selected students were gathered independently on their school grades from three Italian high schools. They were all in their final year of studies, thus being representative of the Italian secondary educational system. No distinction in terms of socio-economic backgrounds was applied in order to guarantee a random sampling representative of the national student population. All of the interviewed researchers had extensive training in STEM disciplines and either possessed or were pursuing a PhD related to complex systems. No selection of researchers based on their backgrounds was actively performed. Both the populations were balanced between male and female participants (see Stella~et~al.~\cite{stella2019forma} for more details).

All participants voluntarily enrolled in the cognitive experiment leading to the construction of FMNs by providing: (i) free associations and (ii) valence scores. The~first type of cognitive data was gathered by means of a continuous free association task~\cite{de2013better}, i.e.,~``write up to three words coming to your mind when reading a given cue'' (for instance, ``complex''). As~already reported in the Introduction, free associations are powerful predictors of a variety of cognitive processes, despite being independent or \textit{free} from a specific definition. A~selection of 50 scientific and educational concepts was used as a set of cues for the free association game. The~associates were then linked to the cue, but~not between themselves, thus forming a network of concepts. In~addition to producing free associations, participants also had to rate their perception of each single word on a Likert scale ranging from 1 (very negative concept) to 5 (very positive concept). A~statistical analysis enabled Stella and colleagues to attribute valence labels ``positive'', ``negative'', and ``neutral'' to each concept, cfr.~\cite{stella2019forma}. 

Given the distinctive combination of associations and valence labels,~the authors in \cite{stella2019forma} extended the psycholinguistic concept of word valence~\cite{smeets2006effect}, i.e.,~how positively/negatively a given word is perceived. Stella and colleagues introduced and tested the network measure of \textit{emotional valence aura}, i.e.,~the tendency for a positive/negative/neutral concept to be surrounded mainly by other positive/negative/neutral concepts. The~authors showed how negative emotional auras are powerful identifiers of anxiety-eliciting concepts, in~agreement with other studies indicating a connection between accumulation of negative word valence and stress prediction~\cite{smeets2006effect}. Positive emotional auras were not found to indicate any specific emotional state but were rather related to positive stances toward a specific topic. This manuscript will assess the emotional auras of educational concepts that were not described or further investigated in previous approaches like~\cite{stella2019forma} or~\cite{stella2019formaindiv}.

\begin{figure}[H]
\centering
\includegraphics[height=7cm]{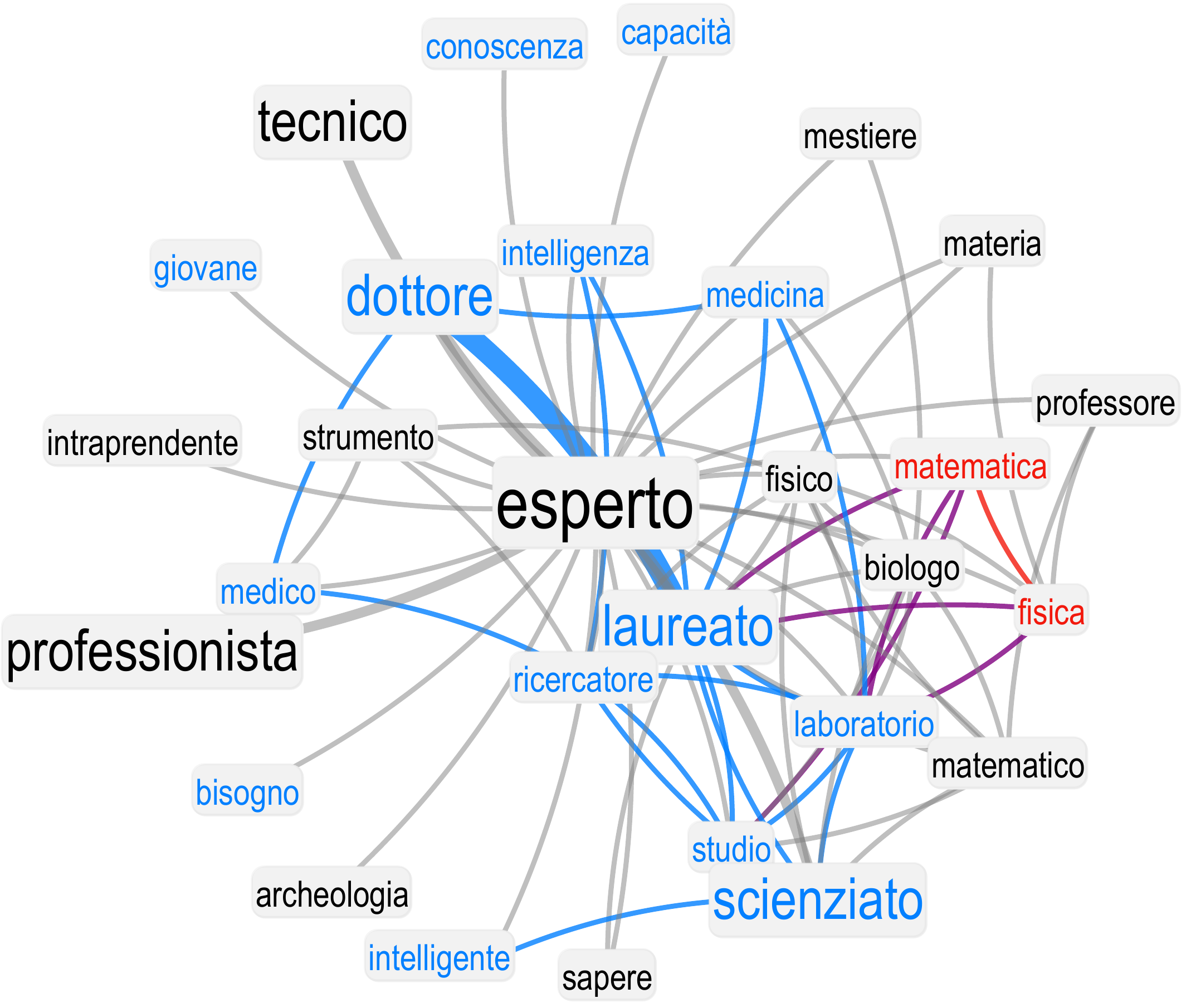}
\caption{Example of a forma mentis network.
 Nodes represent concepts linked by empirical free associations. Conceptual links made by two or more people are thicker. Positive (negative) words are highlighted in cyan (red). Links between positive (negative) words are highlighted with the same color. This example is based on Italian associations made by students relative to the word esperto/expert. {Larger words are also closer to all other associated concepts, i.e.,~on average, there are fewer associations connecting concepts}.}
\label{fig:1}
\end{figure}

An example of forma mentis network is reported in Figure~\ref{fig:1}, where the concepts associated by students to ``esperto'' (expert) are reported {with different font sizes. Larger words are also closer}~\cite{newman2018networks} {to all other associated concepts in the overall forma mentis network, i.e.,~on average there are fewer associations connecting a word of larger font size to all other concepts. In~the reminder, in~denser neighborhoods, a smaller baseline font size was used for visualization purposes, but differences in font sizes were kept. In~Figure~\ref{fig:1}}, notice how ``esperto'' is perceived as a neutral concept (highlighted in black) by students but also surrounded mostly by positive concepts (highlighted in cyan). This~positive aura indicates that, overall, students display a positive stance toward experts in general. This~information would be lost by considering only the valence of ``esperto'' by itself. Notice how the network representation of linked concepts provides additional, microscopic information on the students' stance toward experts. For~instance, the~students' mental representation of experts is strongly mediated by concrete figures like doctors and scientists, professional figures that students can associate to appropriate features of knowledge-sharing, technical skills, and research expertise. These elements characterize a concrete perception, beyond~general stereotypes, and~an appropriate level of awareness about the roles of experts as displayed by~students.

For the sake of an easier understanding of the network visualizations, all Italian words have been translated by using a majority consensus rule of automated translation services (Google Translate, Bing Microsoft Translator, and DeepL). In~other words, the~most frequent translation was chosen for each single Italian word as obtained from the above translation services. Translation was reported only at the level of labelling network visualizations. The~underlying network topology was kept fixed. This might produce the repetition of different Italian words with the same English translation in the following plots, but it is guaranteed to keep fixed the original structure of associations provided by students. Wherever the translation did not reach a consensus, the~original Italian word was kept. The~author, a~native Italian speaker fluent in English, analyzed the quality of the translation for a random subset of 100 Italian words and agreed with $93\%$ of the automated translations, a~percentage that was considered good enough for mere visualization~purposes.

%%%%%%%%%%%%%%%%%%%%%%%%%%%%%%%%%%%%%%%%%%
\section{Results}

The results of this manuscript are reported at three different stages. Firstly, the~general network structure of forma mentis networks is analyzed and discussed as models of knowledge representation. Secondly, the~emotional auras and neighborhoods of educational concepts are discussed and compared across students and researchers. Thirdly, more focus is devoted to understanding differences and analogies about how students and researchers perceive and associate themselves, reporting on potential stereotypes or distorted, anxious~perceptions.

\subsection{The structure of forma mentis networks crucially depends on~hubs}

Before investigating the specific stances of students and researchers, a~network analysis of FMNs can be insightful for understanding how both these groups organized and perceived their conceptual associations and affective patterns within their STEM and educational~knowledge.

Table~\ref{tab1} reports a few key network statistics for both the students' FMN and the experts' FMN. Despite the students' FMN including almost three times as many concepts as the researchers' FMN, both the networks display analogous: {tendency to form triangles of associations} (global clustering coefficient), {tendency for links to connect lowly connected words to highly connected concepts} (assortativity coefficient), {maximum number of associations connecting any two concepts} (network diameter), and { average number of associations linking any two words} (mean network distance). For~a definition of these measures and a brief discussion of them, please see  Appendix \ref{app1} and~\cite{newman2018networks}. 

\begin{table}[H]
\centering
{%
  \caption{Table of network measures for the students' (SFMN) and the researchers' (RFMN) forma mentis networks, together with their random null models, Random S, and Random E, respectively. Null~models preserve connectivity and the number of associates of a word but randomize links (i.e.,~they~are configuration models~\cite{newman2018networks}). The~parentheses indicate standard errors, so that 11 (1) should be read as $11\pm 1$, and~are based on 50 random realizations.} \label{tab1}%
 \resizebox{\textwidth}{!}{ \begin{tabular}{ccccc} 
  \toprule
  \textbf{Measure} &\textbf{ Students' FMN} & \textbf{Experts' FMN} & N\textbf{ull Model (Students)} & \textbf{Null Model (Experts})\\ \midrule
  Concepts & 4483 & 1616 & 4483 & 1616 \\
  Associations & 11728 & 3185 & 11728 & 3185 \\ 
  Clust. Coef. & 0.045 & 0.042 & 0.035~(2) & 0.025~(2) \\ 
  Assor. Coef. & $-$0.34 & $-$0.048 & $-$0.035~(9) & $-$0.026~(9) \\ 
  Diameter & 7 & 10 & 11~(1) & 10~(1) \\ 
  Mean Distance & 4.08 & 4.53 & 3.99~(3) & 4.27~(8) \\ 
  \bottomrule
  \end{tabular}}
}
\end{table}

{In a forma mentis network, every word has a number of links to other concepts, a~number called also degree in network science}~\cite{newman2018networks}. {The distribution of degrees for both the students' and the researchers' FMN is heavy-tailed (see Supplementary Figure~S1 and Appendix), i.e.,~the probability of finding words with large degrees was not exponentially low}. The~heavy-tailed degree distribution, together with the registered degree disassortativity~\cite{van2015examining}, outlines a network structure where constellations of lowly connected words tend to link mainly to hubs. Furthermore, since words do not tend to cluster together, there are fewer different paths connecting them and mainly going through hubs. Are hubs effective in connecting together concepts through a handful of associations? The mean distance between concepts indicates that on average every two concepts are connected by 4 or 5 degrees of separation, i.e.,~conceptual associations. In~a network with thousands of concept, such a short mean distance indicates the crucial role played by hubs in connecting concepts with each other at distances close to those observed in the randomized networks. These results indicate that both the forma mentis networks display a peculiar \textit{small-world structure}, guaranteeing short distances between words like in randomized networks, but~also distributing connectivity mainly through hubs rather than through clustering, differently from the standard definition of small-world networks where high clustering is usually observed, instead (cfr.~\cite{newman2018networks}).

The above results indicate that hub-concepts play an important role in the structured mindsets of both students and researchers. When describing educational concepts, the~rest of the analysis will focus mainly on hubs of relevance for educational~settings.

\subsection{Hubs related to education reveal the stances toward teachers, study, and~school}

\begin{figure}[H]
\centering
\includegraphics[height=7cm]{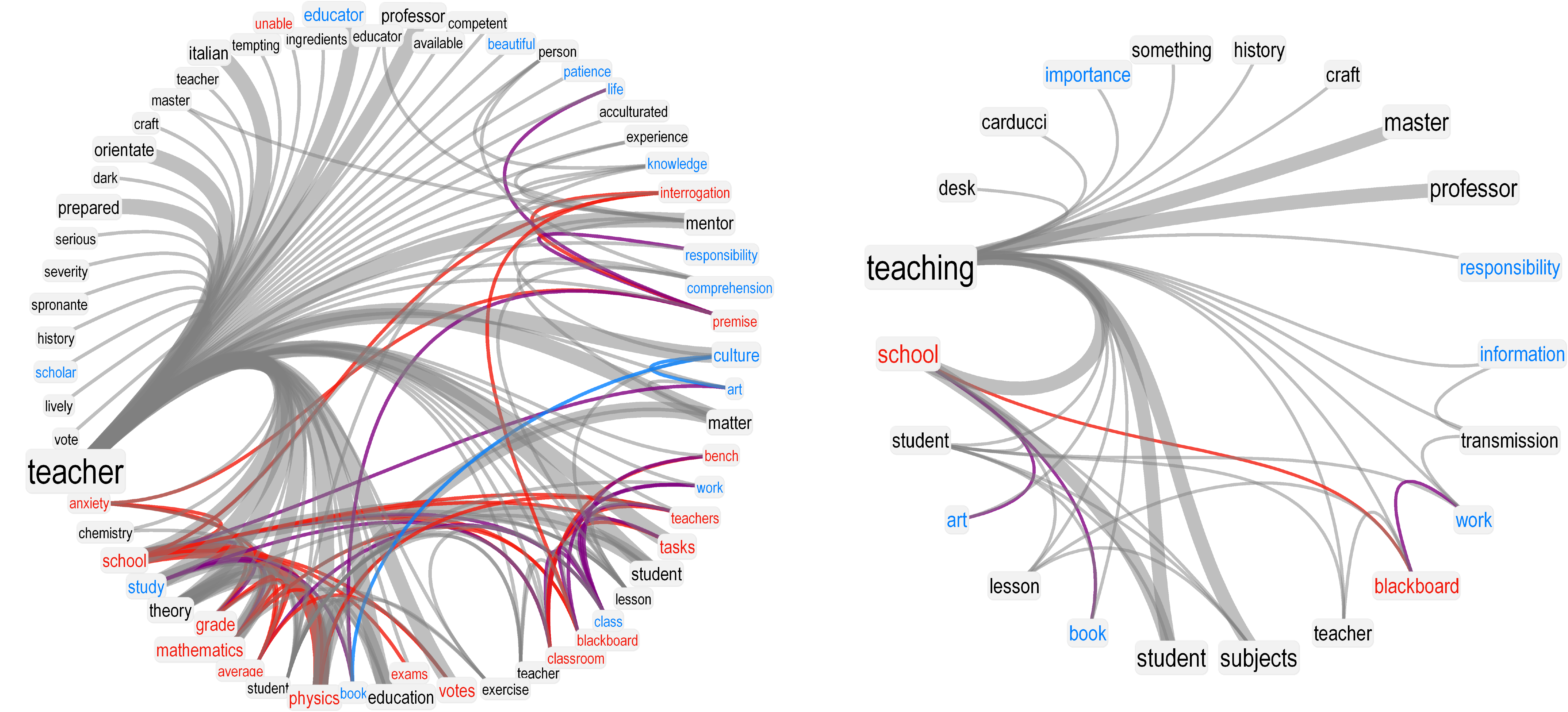}
\includegraphics[height=7cm]{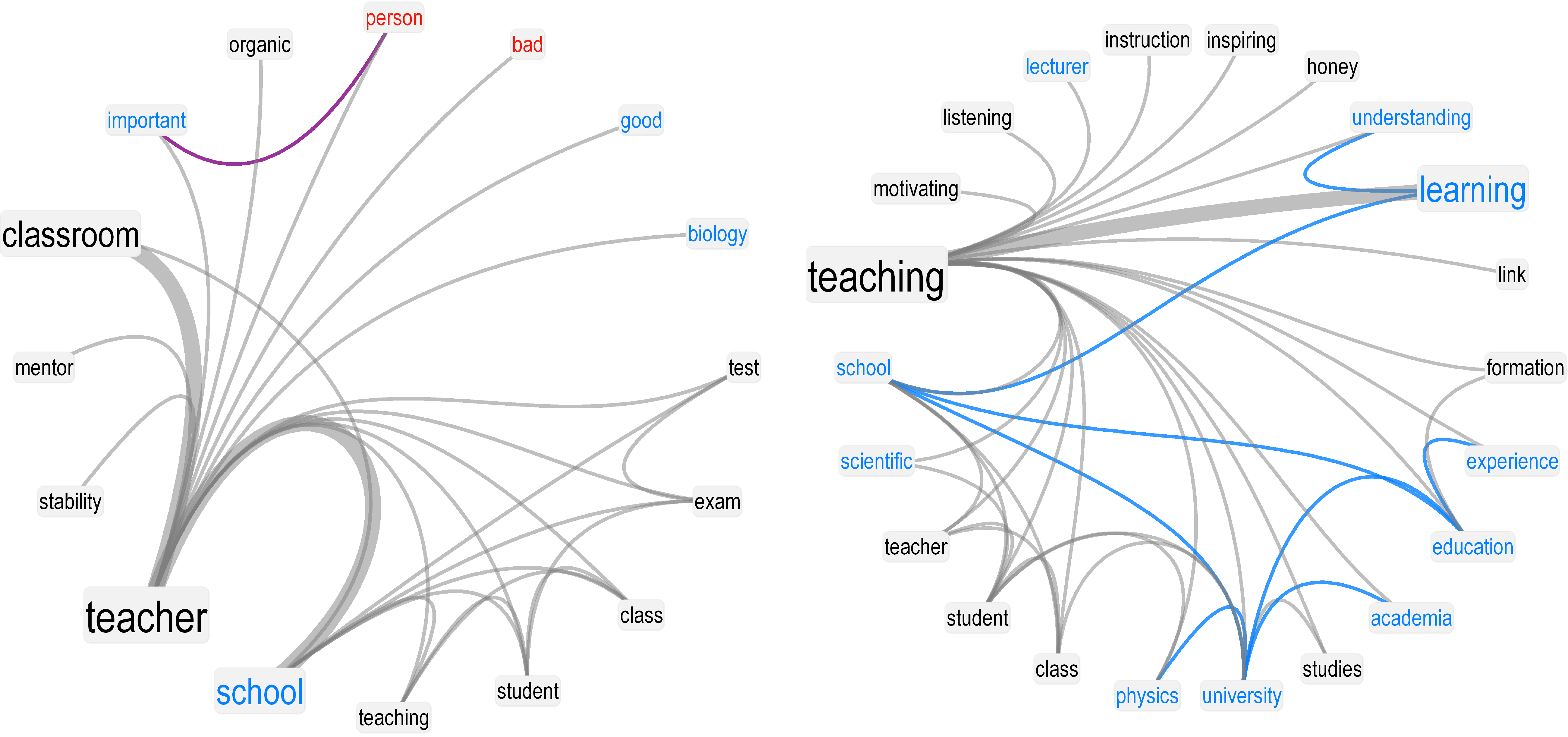}
\caption{\textbf{Above:} Conceptual associations and perceptions for ``teacher'' (left) and ``teaching'' (right) in the students' forma mentis network. Conceptual links made by two or more people are thicker. Positive (negative) words are highlighted in cyan (red). Italian words were translated in English (see Methods).
\textbf{Below:} Conceptual associations and perceptions for ``teacher'' (left) and ``teaching'' (right) in the researchers'~FMN.}
\label{fig:2}
\end{figure}

``Teacher'' is a hub in the forma mentis networks of both students and researchers (see Figure~\ref{fig:2}). Students produced a considerably more clustered neighborhood for ``teacher'' rather than for ``teaching'', indicative of their closer experience with teachers rather than with the act of teaching. Researchers, instead, associated more concepts to ``teaching'' rather than to ``teacher''. Since most of the interviewed researchers had also teaching experience as demonstrators, lecturers, or teaching assistants, it is only expected for them to have more experience with teaching than students and hence provide a richer, more connected mental representation of such concept. For~students, ``teacher'' by itself is a neutral concept, mostly surrounded by neutral concepts. The~presence of a non-trivial fraction of negative and positive concepts associated with ``teacher'' indicates a mixed stance. In~the students' data, most of the negative perception of ``teacher'' revolves around physics and mathematics, which are perceived negatively~\cite{stella2019forma}, but~also toward exams, grades, and quizzes, concepts that include also associations to ``anxiety''. This mental structure of the students' mindset indicates a perceived anxiety tied to the testing and exams that are part of a teacher's duty. However, the~students' perception is not limited by negative concepts, but it rather includes a variety of positively-perceived words such as ``beautiful'', ``patience'', ``knowledge'' and ``culture". Importantly, these words indicate an understanding and appreciation of students toward the educational role of teachers, who are perceived as promoters of knowledge and culture. The~researchers' stance toward teachers is way more pragmatic, focusing mainly on the formal aspects of teaching such as ``students", ``school'', and ``class''. STEM experts perceive ``teaching'' as a neutral concept but surround it with a positive emotional aura, identifying a strong association between ``teaching'' and ``learning'' that is missing in students. While students focus on learning outcomes in terms of grades and appreciate the orienteering figure of teachers, researchers focus more on the places of teaching and on learning itself. Interestingly, the students' perception of ``learning'' is almost completely free from the negative concepts found in the neighborhood of ``teacher", suggesting that students, like researchers, are aware of the difference between the act of teaching and the figures performing such an act in a school~setting.

\begin{figure}[H]
\centering
\includegraphics[height=7cm]{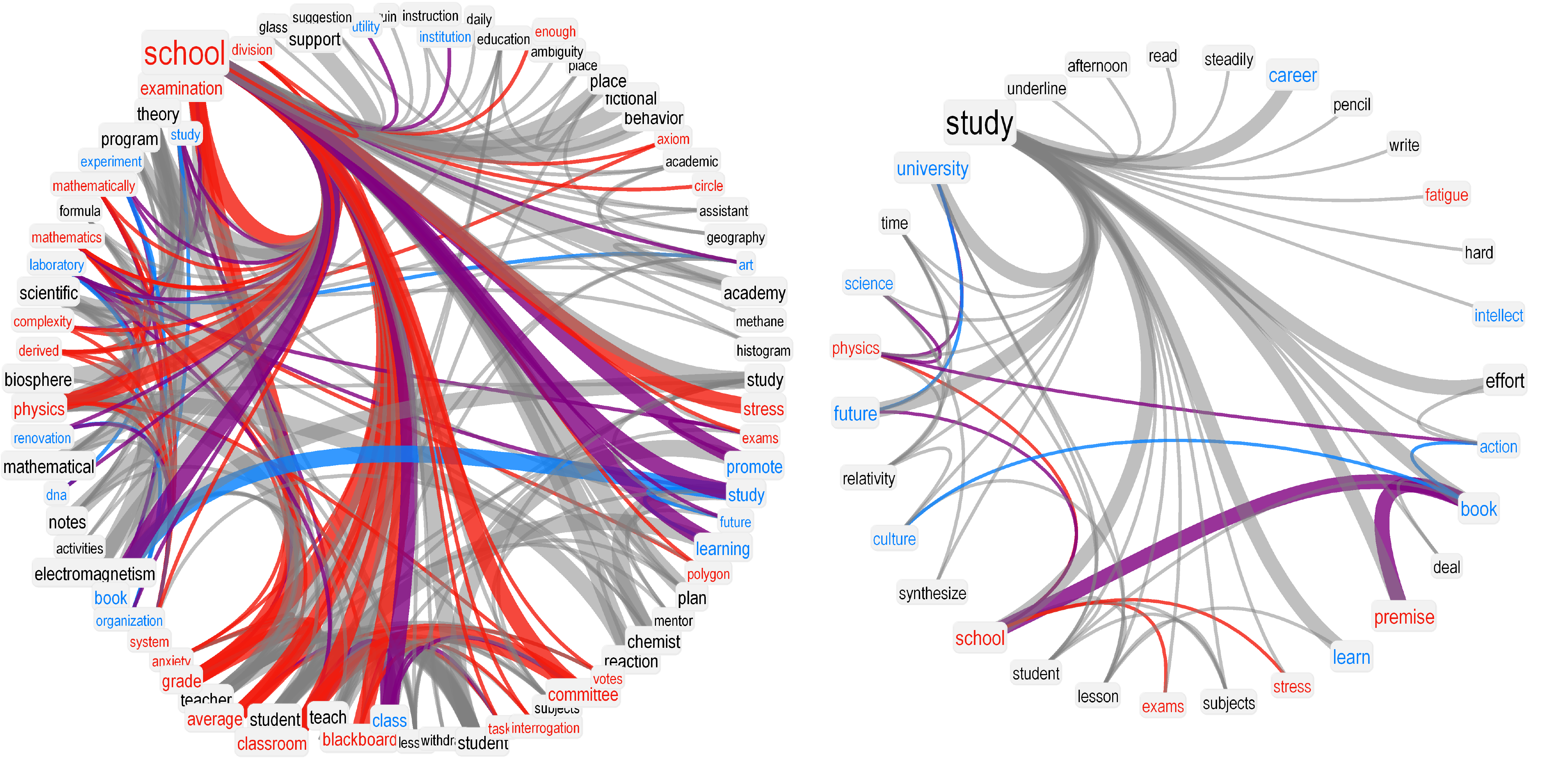}
\includegraphics[height=7cm]{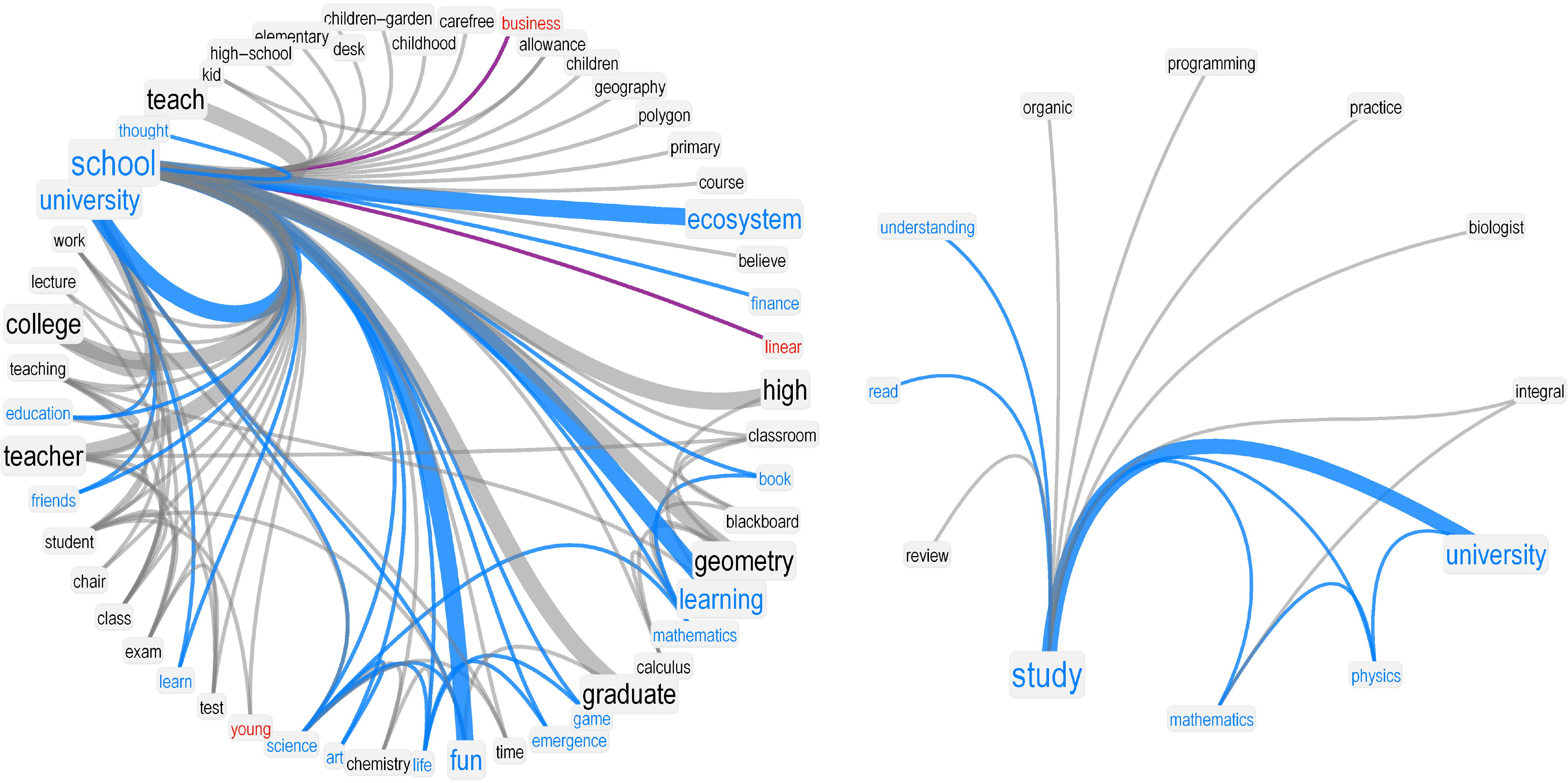}
\caption{\textbf{Above:} {Conceptual associations in the students' FMN and perceptions for ``school'' (left) and ``study'' (right)}. \textbf{Below:} {Conceptual associations in the researchers' FMN and perceptions for ``school'' (left) and ``study'' (right)}. Conceptual links made by two or more people are thicker. Positive (negative) words are highlighted in cyan (red). Italian words were translated in English (see Methods).}
\label{fig:3}
\end{figure}

The differences in how students and researchers perceive teachers motivate a further look at the teachers' educational context. ``Study'' is a hub in the students' forma mentis network. Figure~\ref{fig:3} highlights the stances of ``study'' as perceived by students and researchers. STEM experts perceive ``study'' as a positive concept and associate it with other positive and neutral concepts. These~associations are mainly pragmatic, as~researchers associate ``study'' mainly with ``university'' and other STEM subjects. The~students' mindset around ``study'' is richer in both concrete and abstract concepts. Students perceive ``study'' as a neutral concept connected to positive words related to personal development (e.g., future, culture, career). Even in this stance, a negative outlook on exams and their associations with ``stress'' both persist. Interestingly, one of these negative associations emerges from~``school''.

``School'' is a hub in both the students' and researchers' forma mentis networks. Researchers perceive ``school'' as a positive concept, surrounded by a positive emotional aura (see Figure~\ref{fig:3}). Associations with concepts like ``fun'', ``friends'', ``high-school'', and ``childhood'' suggest that, for researchers, the school is a mental construct from the past and an important socialization avenue. The~researchers' perception of ``school'' is devoid from any anxiety-eliciting associations, which~are present in the students' mindset, instead. The~overall mindset of the 159 interviewed students identifies ``school'' as a negative concept, surrounded by a mostly negative emotional aura. This~negative perception originates mainly from quantitative disciplines such as maths and physics but also from concepts related to assessment like ``committee'', ``exams'', and ``grades''. Even in the stance toward school, anxiety-related associations to ``stress'' and ``anxiety'' are present. The~emerging picture is a generally anxious perception that students have of the school setting, in~particular toward well-known anxiety eliciting STEM disciplines~\cite{mallow2006science,nunez2013effects,siew2019using} and, more importantly, toward the assessment system of grading and tests, which has also been documented in other educational systems~\cite{zeidner2010test}. Even within this negative aura, students are able to identify the important role played by school in promoting learning and education, which are positive concepts associated with ``school''~itself.

\subsection{On the perception of anxiety and~fun}

\begin{figure}[H]
\centering
\includegraphics[width=14cm]{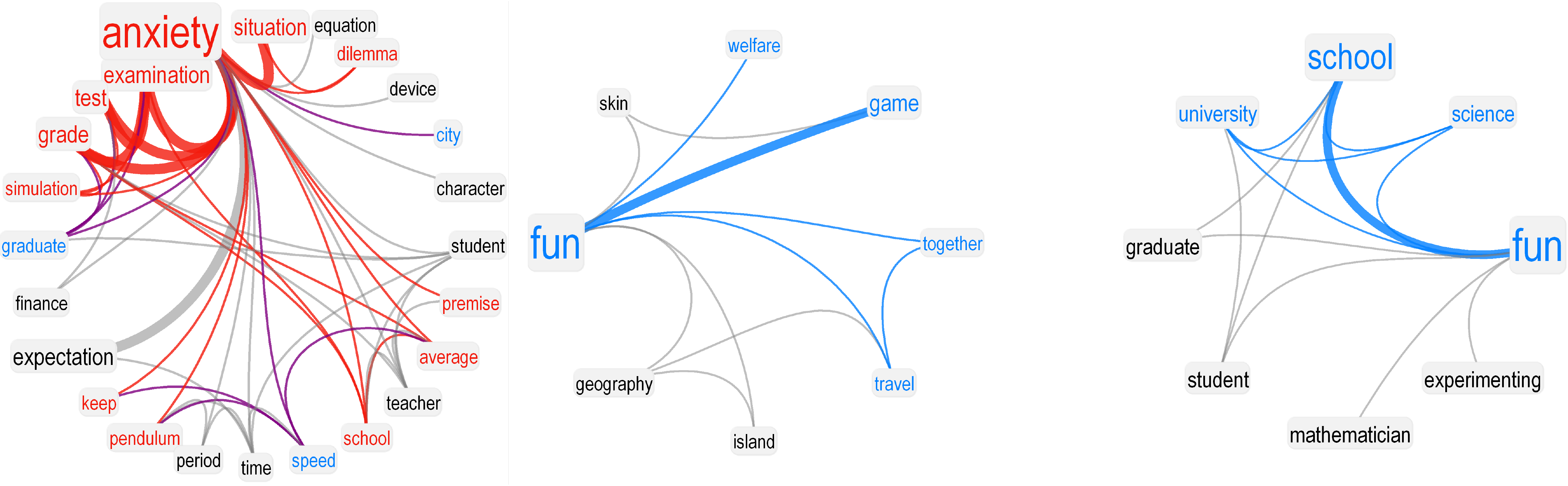}
\caption{\textbf{Left:} Conceptual associations and perceptions for ``anxiety'' and ``fun'' in the students' FMN. \textbf{Right:} Conceptual associations and perceptions for ``fun'' in the researchers' FMN. ``Anxiety'' was not reported or associated by the~researchers.}
\label{fig:4}
\end{figure}

The persistent presence of ``anxiety'' across the students' perceptions of ``school'', ``study'', and ``teacher'' indicates a conceptual relationship, present in the students' mindset, between~educational settings, actions, and actors and stress. A~previous approach with forma mentis networks identified anxiety coming from a distorted perception of STEM subjects~\cite{stella2019forma}, but it did not focus on the mental representation of ``anxiety'' itself, which is reported in Figure~\ref{fig:4}. It is important to underline that ``anxiety'' was not among the selected cues, but it was rather a target word associated by students. Researchers did not provide such association when tested with the same set of cues. Confirming the relationship between negative valence, negative associations, and anxiety eliciting found in~\cite{smeets2006effect} and in~\cite{stella2019forma}, ``anxiety'' is a negative concept surrounded by a negative aura in the students' FMN. Students provided strong conceptual associations between ``anxiety'' and a set of educational concepts related to personal assessment, e.g.,~test, grade, simulation, examination, school. The~strong link with ``expectation'' suggests that the test anxiety perceived by students is related to biased expectations, possibly in relation to performance in exams and tests. Notice that no associations are made between ``anxiety'' and ``physics'' or ``maths'', so that the anxiety surrounding these STEM disciplines found in~\cite{stella2019forma} should have a different nature in comparison to the test anxiety detected~here.

``Fun'' can be considered as antithetic to ``anxiety'', and it is associated differently between the students' and the researchers' FMNs, whereas students associate ``fun'' mainly with recreational activities, and researchers provide connecting fun with making science, studying, and~learning. 
   
\unskip

\subsection{Hubs related to self-perceptions reveal positive, mostly non-stereotypical attitudes toward scientists and~students}

\begin{figure}[H]
\centering
\includegraphics[width=14cm]{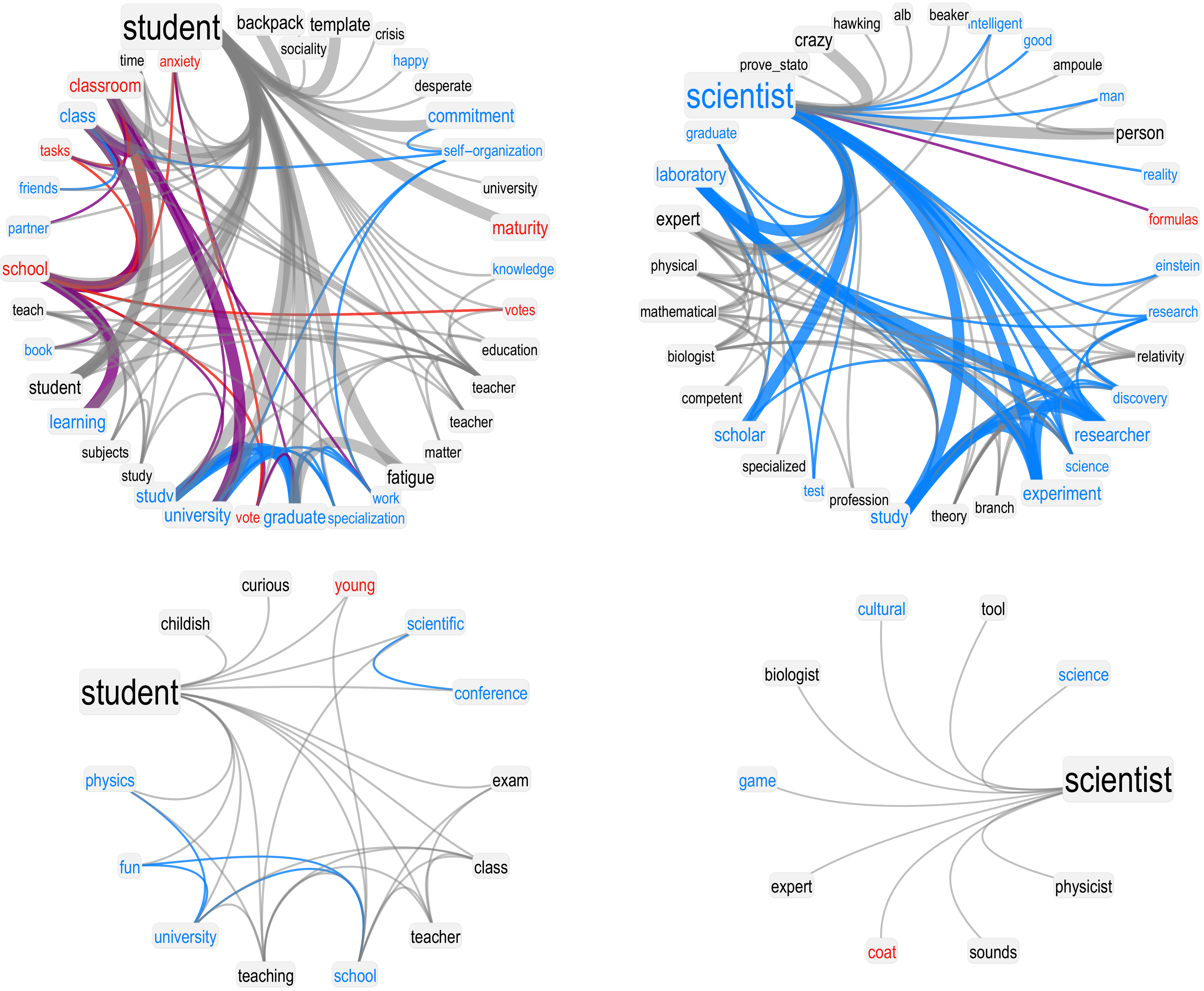}
\caption{\textbf{Above:} Conceptual associations and perceptions for ``student'' and ``scientist'' in the students' FMN. \textbf{Below:} neighborhoods for ``student'' and ``scientist'' and in the researchers' FMN. Conceptual links made by two or more people are thicker. Positive (negative) words are highlighted in cyan (red). Italian words were translated in English (see Methods).}
\label{fig:5}
\end{figure}

Beyond their perception of surrounding reality, it is interesting to investigate also how students and young scientists perceive themselves. These self-perceptions can be informative about the presence of specific stereotypes or potential biases about the main actors of STEM learning and research. Figure~\ref{fig:5} reports the FMN structure around ``student'' and ``scientist'' in both high schoolers and researchers. Students build up a mental representation of themselves that is neutral but associated with several positive concepts about personal growth, education and career building, as~indicated by links with ``happy'', ``commitment'', ``learning'' and ``career''. Free associations and affect labels identify a cluster of concepts linked to students that is about ``work'' and ``specialization'', suggesting the students' projections about their future in the job market. Even in the positive aura of ``student'' as perceived by students, there are negative outliers related to school, grades, and anxiety. ``Maturity'' in Italian indicates also the final exam at the end of high school, a~concept perceived negatively and strongly related by students to their mental construct or self-perception. In addition, researchers perceive the idea of ``student'' as a neutral concept and surround it with positive words. In~the researchers' FMN, however, students are represented not only as components of the educational system but also as active agents of the research environment (see associations with ``scientific'' and ``conference''). Differently from students, researchers do not associate ``student'' to any concept specifically related to the job market. The~researchers' perception of ``scientist'' is neutral, with~few associations mainly focused to specific types of scientists. The~students' FMN offers a much richer stance. Students perceive ``scientist'' as a positive concept surrounded by a positive emotional aura. This finding, in~addition to the specific associations provided like ``scientist-good'' or ``scientist-intelligent'', indicate an appreciation of students toward scientists in general. Through clusters of associations with research and STEM disciplines, students display a good awareness of the role played by scientists in promoting science across the spectrum of STEM disciplines through ``research'', ``theory'' and ``experiments''. Within~this overwhelmingly positive, detailed and non-stereotypical perception of scientists, there is also a strong cognitive association between ``scientist'' and ``crazy'', indicating the permanence of a small stereotype about ``mad scientists''. The~forma mentis network of students indicates that the ``mad genius'' stereotype co-exists with a positive  and mostly non-stereotypical perception of scientists. The~ability for the forma mentis network to capture potential stereotypes in the mental representation of specific categories of individuals opens important directions for future~research.

\unskip

\section{Discussion}

This manuscript adopted the recent framework of forma mentis network~\cite{stella2019forma} for reconstructing, quantifying and comparing how high school students and researchers perceive actors, places, and elements of relevance for education. The~analysis focused on concepts such as ``teacher'', ``study'', ``learning'', ``school'', ``student'', and ``researcher'' and identified several differences in the mindsets of students and researchers. Consistently across all the stances focusing on such concepts, STEM experts reported rather pragmatic views, more focused on STEM disciplines and related to learning as a positive, fun, and rewarding experience. Instead, students exhibited a more complex range of perceptions, ranging from the negative associations with STEM disciplines like statistics or physics or maths to the positive identification of ``study'', ``teachers'', and ``scientists'' as promoters of personal and societal development through knowledge and education. In~comparison to STEM experts, students exhibited a stronger awareness of the relevance of education for better approaching the job~market.

The most crucial difference in mindsets was found within the mental representation of ``school'', which is perceived and linked negatively by students and positively by researchers. Considering that the current investigation involves 159 students, this difference has to be pinpointed to systematic differences in the perception of school rather than in idiosyncratic differences attributed to personal preferences. Considering the presence of several strong conceptual associations between ``school'', ``stress'', ``anxiety'', and other concepts related to learning assessment (e.g.,~``grades'', ``committee'', ``exams''), forma mentis networks highlight the presence of an anxious perception of school affecting the whole student population. Anxiety-focused conceptual associations persisted in almost every other educational stance investigated here, indicating a strong influence of such negative emotion over the perception of education as portrayed by students. On~the one hand, it might be that this anxiety is only the symptom of the final exam (''maturità'' in Italian), at the end of high school, all the interviewed students were preparing for. On~the other hand, this anxiety did not interest all STEM disciplines or educational cues in the FMN (see also~\cite{stella2019forma}) but rather focused only on specific topics, thus suggesting that the students' anxiety did not distribute on their whole scientific mindset but rather concentrated only around specific parts of it. Such concentration is an indication that there might be not just one general feeling of anxiety permeating all the tested mindsets but rather \textit{multiple} types of anxiety affecting mental constructs in different ways. Detecting and understanding these different sources for anxiety becomes key to enabling intervention policies for reducing stress in students. In~their FMN, students portrayed ``anxiety'' as strongly connected to performance anxiety in tests, a~well documented cause of stress~\cite{cassady2002cognitive,zeidner2010test} that could be prevented by promoting creative interactions through cooperative learning of STEM subjects, as~tested in chemistry teaching classes~\cite{oludipe2010effect}. Conceptual associations between ``anxiety'' and STEM subjects were missing in the students' FMN, despite anxiety-eliciting being detected in the negative auras of disciplines like maths, physics, and statistics~\cite{stella2019forma}. These missing links suggest the presence of a different type of anxiety affecting how students perceive these subjects, beyond~assessment in itself. Importantly, several studies identified STEM subjects as potential sources of distress in students~\cite{mallow2006science,nunez2013effects,valenti2016adolescents,siew2019using}, often caused by an erroneous perception that these disciplines are inherently hard, require too much effort, cannot be understood by anyone and are dry, without~any specific relevance for everyday life. This distorted perception is partly due also to specific teaching styles presenting STEM subjects as separate compartments incomplete by themselves and detached from reality~\cite{van2018educate,bruun2017network}. These issues might be tackled by using innovative tools from cognitive network science, mapping students' interactions, discussions, mindsets, and performance~\cite{koponen2014systemic,bruun2016networks,tyumeneva2017distinctive,siew2019using,stella2019forma} and offering data-driven support to managing teaching and learning in classroom settings~\cite{mika2019maths}. In addition, {the adoption of persuasive technologies}~\cite{ahmad2018study} and
a renewed focus on Jantsch's interdisciplinarity~\cite{jantsch1972inter,van2018educate,cramer2018network} could help with addressing the above distorted perception by making students aware of the creative, fun side of the above STEM subjects in relation to the complexity of the real world. The~current analysis indicates that such connection between ``fun'' and ``science'' characterises STEM experts while it is missing in students. As~a concrete example, building conceptual links between history and physics and presenting students with reconstructions about the historical role of physics experiments of the 19th century has been indicated as a rather powerful way of favoring physics teaching~\cite{koponen2006generative}. Although~resistance-to-change~\cite{tsiouplis2019rethinking} poses a challenge for innovating teaching styles and curricula, the~advent of tools reducing anxiety and favoring concentration in students is important not only for its self-evident psychological benefits but also in terms of facilitating and enhancing students' performance in STEM~\cite{bodin2012role,lehtamo2018connection} and students' appreciation of STEM careers~\cite{valenti2016adolescents}.

The microscopic structure of students' and researchers' mindsets reconstructed by FMNs enabled also a comparison between self-perceptions. Young scientists reported a schematic perception of themselves, mainly in relation to STEM fields. They also perceived students in relation to the educational environment but also as associated with scientific conferences. Students perceived themselves as mainly neutral, with~a positive attitude toward personal growth through learning, work and university education. Importantly, students perceived scientists as an overwhelmingly positive figure, surrounded by a positive emotional aura and associated with traits of goodness and intelligence and devoid of any anxiety-eliciting conceptual association. High schoolers also associated ``scientist'' with ``crazy'', suggesting some awareness of the stereotypical mental picture of the mad genius~\cite{finson1995development,rahm2002scientist,haynes2016whatever}, a~stereotype overwhelmingly present in fiction and media and which dangerously obfuscates the work of scientists and the scientific method. The~ability for students to associate ``scientist'' both with general-level concepts about knowledge and specific tools and fields of research in STEM indicates a mainly non-stereotypical perception, which is an important stepping stone for improving their perception of STEM subjects too. {Importantly, the~overall non-stereotypical perception of scientists reported by students indicates that stereotype threat}~\cite{shapiro2012role} {did not affect the forma mentis of students and cannot, therefore, be considered a strong cause for the negative, anxious perception reported in}~\cite{stella2019forma}. Overall, these findings pose forma mentis networks as a novel alternative to other techniques quantifying scientists' perception such as picture drawing or questionnaires~\cite{finson1995development} for future research~directions.

Although forma mentis networks offer important information about the mental representation of stances, the~above analysis is limited under some aspects. For~instance, the~current analysis focuses on the population level but does not provide insights about individual students or researchers. Although~further research in this direction is still necessary, a~first attempt of using individual-level forma mentis networks has been done by Stella and Zaytseva, who successfully used FMNs for detecting changes in the mindsets of a small group of students over a summer job experience~\cite{stella2019formaindiv}. Building individual-level FMNs requires more cues and is therefore more time consuming when interviewing large groups of participants, but it can provide longitudinal characterization of stances in specific populations. Another limit of the current analysis is the lack of a temporal dimension. The~currently analyzed snapshots enable distinguishing crucial differences between students and experts but cannot pinpoint exactly how these differences emerged over time. Following a cohort of students over a short amount of time, analogously to Stella and Zaytseva's work~\cite{stella2019formaindiv}, might better identify the role played by competence acquisition over the detected students'~anxiety. 

\section{Conclusions} 

{Beyond the above data-related limitations, forma mentis networks have the power to explore how different stances co-exist together in the overall mindset of a given group. In~this way, FMNs~can be used for detecting different types of anxiety affecting the perception of STEM within student populations, without~the need for hard-coding specific questions as in standard surveys. As~reported in the current analysis, within~the same framework of FMNs, evidence for STEM anxiety}~\cite{mallow2006science}, {test~anxiety}~\cite{zeidner2010test}, and {a non-stereotypical perception of scientists, indicating a lack of stereotype threat patterns}~\cite{shapiro2012role}, {were all identified as co-occurring within the mindset of high-school students}. 

Overall, the~simplicity of the cognitive task behind network construction, the~microscopic access to conceptual knowledge and emotions and the quantitative results reported in the above analysis indicate forma mentis networks as providers of important information for better understanding and addressing anxiety and stereotypical perceptions in educational~contexts and intervention policies.

%\begin{figure}[ht!]
%\centering
%\includegraphics[height=5cm]{ExampleFormaMentisNetwork.pdf}
%\caption{Cumulative degree distributions for the students' and the researchers' FMNs, identifying the probability $P(X)$ of finding a word with at least $X$ associates.}
%\label{figsup:1}
%\end{figure}
\vspace{6pt}

%%%%%%%%%%%%%%%%%%%%%%%%%%%%%%%%%%%%%%%%%%
\section*{Funding}
{This research received no external~funding.}

%%%%%%%%%%%%%%%%%%%%%%%%%%%%%%%%%%%%%%%%%%
\section*{Acknowledgements}
{The author acknowledges Ruth Lazkoz for interesting discussion about the stereotypes of scientists in the relevant literature. The~author acknowledges also Sarah De Nigris, Aleksandra Aloric, and Cynthia S. Q. Siew for insightful discussion about education and network~science.}

%%%%%%%%%%%%%%%%%%%%%%%%%%%%%%%%%%%%%%%%%%
\section*{Conflicts of Interest}
{The author is employed at Complex Science Consulting. The~funders had no role in the design of the study; in the collection, analyses, or~interpretation of data; in the writing of the manuscript, or~in the decision to publish the~results.}

\section*{Appendix 1: Network Measures of Forma Mentis Networks}\label{app1}

In the main text, several network metrics for the FMNs of students and researchers are presented and briefly discussed. This Appendix provides additional details about such metrics, focusing over the global clustering coefficient reported in Table~\ref{tab1} and the degree distribution discussed in the Results~section.

Global clustering coefficient measures how many triangles are present in the network~\cite{newman2018networks} and can range between 0 (e.g., tree network, no triangles) and 1 (e.g., complete graph, all possible triangles are formed). The~modest values of clustering detected in both of the FMNs indicate a weak tendency for concepts to form triangles of associations and cluster together. Randomizing conceptual associations while preserving the number of links of each word (i.e.,~degree) produces null models with lower clustering coefficients but of the same order of magnitude of the empirical networks. This comparison indicates that the degree distribution partially induces but cannot fully explain word clustering. Additional cognitive phenomena, such as semantic similarity~\cite{van2015examining}, might foster triangle formation and global clustering despite network construction (where recalled words are linked only to the cue and not between themselves).

A closer look at the degree distribution, reported in Supplementary Figure~S1, indicates that both the forma mentis networks are rich in low-degree nodes and feature heavy-tailed degree distributions with hubs~\cite{newman2018networks}, i.e.,~a few words involved in a large fraction of associations. The~tipping points in the degree distribution are used for identifying hubs, i.e.,~nodes with degree higher than 30 (13) in the students' (researchers') FMN. 

%\externalbibliography{yes}
%\bibliography{bibbi.bib}

\bibliographystyle{plain}

%%%%%%%%%%%%%%%%%%%%%%%%%%%%%%%%%%%%%%%%%%
\end{document}